\documentclass[manuscript]{aastex}
\usepackage{CJK}

\shorttitle{THE $^{13}$C($\alpha$,\,$n$)$^{16}$O REACTION AND AGB
STARS} \shortauthors{GUO et al.}

\begin{document}
\begin{CJK*}{GBK}{song}
\title{New determination of the $^{13}$C($\alpha$,\,$n$)$^{16}$O reaction
rate and its influence on the $s$-process nucleosynthesis in AGB
stars}

\author{B. Guo\altaffilmark{1,6}, Z. H. Li\altaffilmark{1},
M. Lugaro\altaffilmark{2}, J. Buntain\altaffilmark{2}, D. Y.
Pang\altaffilmark{3,7}, Y. J. Li\altaffilmark{1}, J.
Su\altaffilmark{1}, S. Q. Yan\altaffilmark{1}, X. X.
Bai\altaffilmark{1}, Y. S. Chen\altaffilmark{1}, Q. W.
Fan\altaffilmark{1}, S. J. Jin\altaffilmark{1}, A. I.
Karakas\altaffilmark{4}, E. T. Li\altaffilmark{1}, Z. C.
Li\altaffilmark{1}, G. Lian\altaffilmark{1}, J. C.
Liu\altaffilmark{1}, X. Liu\altaffilmark{1}, J. R.
Shi\altaffilmark{5}, N. C. Shu\altaffilmark{1}, B. X.
Wang\altaffilmark{1}, Y. B. Wang\altaffilmark{1}, S.
Zeng\altaffilmark{1} and W. P. Liu\altaffilmark{1,6}}

\affil{\altaffilmark{1}China Institute of Atomic Energy, P.O. Box
275(1), Beijing 102413, China}

\affil{\altaffilmark{2}Monash Centre for Astrophysics, Monash
University, Clayton 3800, Victoria, Australia}

\affil{\altaffilmark{3}School of Physics and State Key Laboratory of
Nuclear Physics and Technology, Peking University, Beijing 100871,
China}

\affil{\altaffilmark{4}Research School of Astronomy \& Astrophysics,
Mount Stromlo Observatory, Weston Creek ACT 2611, Australia}

\affil{\altaffilmark{5}National Astronomical Observatories, Chinese
Academy of Science, Beijing 100012, China}

\altaffiltext{6}{Electronic address: wpliu@ciae.ac.cn,
guobing@ciae.ac.cn}
\altaffiltext{7}{Current address: School of
Physics and Nuclear Energy Engineering, Beihang University}

\begin{abstract}
We present a new measurement of the $\alpha$-spectroscopic factor
($S_\alpha$) and the asymptotic normalization coefficient (ANC) for
the 6.356 MeV 1/2$^+$ subthreshold state of $^{17}$O through the
$^{13}$C($^{11}$B,\,$^{7}$Li)$^{17}$O transfer reaction and we
determine the $\alpha$-width of this state. This is believed to have
a strong effect on the rate of the $^{13}$C($\alpha$,\,$n$)$^{16}$O
reaction, the main neutron source for {\it slow} neutron captures
(the $s$-process) in asymptotic giant branch (AGB) stars. Based on
the new width we derive the astrophysical S-factor and the stellar
rate of the $^{13}$C($\alpha$,\,$n$)$^{16}$O reaction. At a
temperature of 100 MK our rate is roughly two times larger than that
by \citet{cau88} and two times smaller than that recommended by the
NACRE compilation. We use the new rate and different rates available
in the literature as input in simulations of AGB stars to study
their influence on the abundances of selected $s$-process elements
and isotopic ratios. There are no changes in the final results using
the different rates for the $^{13}$C($\alpha$,\,$n$)$^{16}$O
reaction when the $^{13}$C burns completely in radiative conditions.
When the $^{13}$C burns in convective conditions, as in stars of
initial mass lower than $\sim$2 $M_\sun$ and in post-AGB stars, some
changes are to be expected, e.g., of up to 25\% for Pb in our
models. These variations will have to be carefully analyzed when
more accurate stellar mixing models and more precise observational
constraints are available.
\end{abstract}


\section{Introduction}

Approximately half of the elements heavier than iron in the universe
are produced via a series of $slow$ neutron capture reactions and
competing $\beta$-decays (the $s$-process). During the $s$-process,
the neutron number density is relatively low, of the order of 10$^7$
n cm$^{-3}$. When the flux reaches an unstable nucleus, it
typically decays rather than capture another neutron and
the $s$-process proceeds via the
isotopes around the valley of $\beta$-stability \citep[e.g.,][]{bur57}.
The astrophysical
sites of the $s$-process are core He and shell C burning in massive
stars for the elements lighter than Sr \citep{pig10}, and the ``He
intershell'' of asymptotic giant branch (AGB) stars for the elements
between Sr and Bi \citep{gal98}. Stars with initial masses lower
than roughly 9 $M_\sun$ reach the AGB phase in the final phases of
their evolution, when both H and He have been exhausted in the core
leaving C and O in electron degenerate conditions. Production of
nuclear energy occurs in the H and He shells, which are located
between the core and the extended convective envelope and are
separated by the thin He intershell layer. AGB stars experience
thermal pulses (TPs) when the usually dormant He burning shell is
suddenly activated. A large amount of energy is released, which
drives convection in the He intershell. During TPs the star expands and cools
and the H burning shell is inactive.
While He burning turns from the convective to the radiative regime,
and eventually switches off, the convective envelope can penetrate
the underlying He intershell and carry to the surface the products
of He burning, in particular carbon and the elements
heavier than iron made by the $s$-process. This mixing process is
known as the ``third dredge-up'' (TDU). After the TDU is ended, the
star contracts and heats up again and H burning resumes until
another TP occurs and the cycle is repeated. This sequence of events
can occur from a few times to hundreds of times, depending on the
stellar mass and the mass-loss rate. AGB stars suffer
from very strong stellar winds, which erode the envelope roughly
within a million years and shed the newly synthesized material mixed
to the surface by the TDU into the interstellar medium. Eventually,
the C-O degenerate core is left as a cooling white dwarf
\citep[see][for a review on AGB stars]{her05}.

According to the current standard model
\citep{gal98,bus99,gor00,lug03a,cri09a}, some protons must diffuse
from the convective envelope into the He intershell at the end of
each TDU in order to produce enough $^{13}$C to account for the
observed abundances of the $s$-process elements at the surface of
AGB stars \citep[see also][]{bus01}. A thin layer is then produced,
known as the $^{13}$C ``pocket'', which is rich in $^{13}$C made via
$^{12}$C($p$,\,$\gamma$)$^{13}$N($\beta^+\nu$)$^{13}$C. When in this
region the temperature reaches about $9 \times 10^7$ K, the
$^{13}$C($\alpha$,\,$n$)$^{16}$O reaction is activated and generates
neutrons that trigger the $s$-process
\citep{hol88,gal88,kap90,kap11}.

Considerable effort has been devoted to the direct measurement of
the $^{13}$C($\alpha$,\,$n$)$^{16}$O cross section
\citep{sek67,dav68,bai73,kel89,dro93,bru93,har05,hei08}. These
measurements have been performed at energies down to 270 keV,
whereas the Gamow window is at 190 $\pm$ 40 keV, corresponding to a
temperature of 100 MK. Since this energy is far below the Coulomb
barrier, the reaction cross section is extremely small and direct
measurement is sensitively limited by background signals and very
difficult to perform in laboratories on the Earth's surface. While a
measurement has been proposed at the underground laboratory of LUNA
\citep{cos09}, at present, the experimental cross sections have to
be extrapolated below 270 keV. A microscopic cluster model analysis
of the $^{13}$C($\alpha$,\,$n$)$^{16}$O and
$^{13}$C($\alpha$,\,$\alpha$) reactions by \citet{des87} suggested
that this extrapolation is critically affected by the 1/2$^+$
subthreshold resonance in $^{17}$O ($E_x$ = 6.356 MeV, just 3 keV
below the $\alpha$-decay threshold). The contribution from this
resonance depends strongly on the $\alpha$-width of the 1/2$^+$
state in $^{17}$O, which can be derived from the spectroscopic
factor ($S_\alpha$) or the asymptotic normalization coefficient
(ANC) of $\alpha$-cluster in this state.

The $S_\alpha$ and the ANC can be determined from the angular
distribution of the direct $\alpha$-transfer reaction using
distorted wave Born approximation (DWBA) or coupled reaction
channels (CRC) analysis. Although three indirect measurements via
the ($^{6}$Li,\,$d$) or the ($^{7}$Li,\,$t$) system have been
performed by \citet{kub03}, \citet{joh06}, and \citet{pel08} to
study the $S_\alpha$ or the ANC of the 1/2$^+$ state, there still
exists a significant discrepancy of up to a factor of $\sim$30 in
the derived $S_\alpha$ and ANC. Therefore, it is interesting to
perform a new measurement of the $S_\alpha$ and the ANC via an
independent transfer reaction. In addition, it is necessary to
understand the impact of the different resulting
$^{13}$C($\alpha$,\,$n$)$^{16}$O rates on the $s$-process
nucleosynthesis in AGB stars.

In this paper we determine a new stellar rate of the
$^{13}$C($\alpha$,\,$n$)$^{16}$O reaction and incorporate it
in calculations of the $s$-process nucleosynthesis in AGB
stars. First, we measure the angular distribution of the
$^{13}$C($^{11}$B,\,$^{7}$Li)$^{17}$O reaction to determine the
$S_\alpha$ and the ANC for the 1/2$^+$ state in $^{17}$O. Using this
experimental ANC we derive the $\alpha$-width for the 1/2$^+$
subthreshold resonance, which is currently the most uncertain
parameter for determining the $^{13}$C($\alpha$,\,$n$)$^{16}$O rate.
Finally, we use the new rate and different rates available in the
literature as input for simulations of AGB stars to study
their influence on the abundance of some selected $s$-process
elements and isotopic ratios.

\section{Measurement and evaluation of the ANC}

\subsection{Experimental procedure}

The measurement of the angular distribution for the
$^{13}$C($^{11}$B,\,$^{7}$Li)$^{17}$O reaction was performed at the
HI-13 tandem accelerator of China Institute of Atomic Energy (CIAE)
in Beijing. We used a $^{11}$B beam with an energy of 50 MeV leading
to the production of the excited states in $^{17}$O at $E_x$ = 3.055
MeV, 3.843 MeV, 4.554 MeV and 6.356 MeV. The angular distribution of
the elastic scattering for the entrance channel
($^{11}$B\,+\,$^{13}$C) was also measured. We used a self-supporting
$^{13}$C target with a thickness of 75\,$\pm$\,6 $\mu$g/cm$^2$ and
an initial purity of 88\%. A 26 MeV $^7$Li beam was also delivered
for measurement of the exit channel ($^7$Li\,+\,$^{17}$O) elastic
scattering. Natural silicon monoxide of 86\,$\pm$\,7 $\mu$g/cm$^2$
was prepared onto a 40\,$\pm$\,3 $\mu$g/cm$^2$ carbon foil, serving
as the oxygen target. In addition, a self-supporting $^{12}$C target
of 66\,$\pm$\,5 $\mu$g/cm$^2$ was used for calibration of the focal
plane and background subtraction during the whole experiment.

To monitor the possible buildup of $^{12}$C, the $^{11}$B elastic
scattering on the $^{13}$C target was measured at the start and at
the end of the measurement for each angle. This showed that the
$^{12}$C buildup was negligible compared to the initial impurity in
the $^{13}$C target, possibly due to the rather low gas pressure
($\leq$\,10$^{-6}$ mb) in the reaction chamber. To determine the
absolute amount of $^{12}$C in the $^{13}$C target we measured the
angular distributions of the $^{11}$B\,+\,$^{12}$C elastic
scattering with both the natural $^{12}$C target and the enriched
$^{13}$C target. The absolute amounts of $^{12}$C and $^{13}$C in
the $^{13}$C target were found to be 9.0\,$\pm$\,0.7 and
66\,$\pm$\,5 $\mu$g/cm$^2$, respectively.

The beam current was measured by a Faraday cup covering an angular
range of $\pm\,$6$^\circ$ in a laboratory frame and used for the
absolute normalization of the cross sections at
$\theta_\mathrm{lab}\,>\,6^\circ$. The Faraday cup was removed when
measuring the cross sections at
$\theta_\mathrm{lab}\,\leq\,6^\circ$. A Si $\Delta E\,-\,E$
telescope located at $\theta_\mathrm{lab}\,=\,25^\circ$ was employed
for the relative normalization of the cross sections at
$\theta_\mathrm{lab}\,\leq\,6^\circ$ by measuring the elastic
scattering of the incident ions on the targets. In addition, the
ratio of current integration in the Faraday cup to the elastic
scattering events was measured at the start and the end of the
measurement for each angle $\theta_\mathrm{lab}\,\leq\,6^\circ$ by
restoring the Faraday cup. The ratios changed by less than 2\%,
which thus led to a reliable normalization of the cross sections at
$\theta_\mathrm{lab}\,\leq\,6^\circ$.

The reaction products were focused and separated by a Q3D magnetic
spectrograph and recorded by a two-dimensional position-sensitive
silicon detector (PSSD, 50\,$\times$\,50 mm) fixed at the focal
plane of the spectrograph. The two-dimensional position information
from the PSSD enabled the products emitted into the acceptable solid
angle to be recorded completely. The energy information from the
PSSD was used to remove the impurities with the same magnetic
rigidity.

\begin{figure}
\epsscale{.80} \plotone{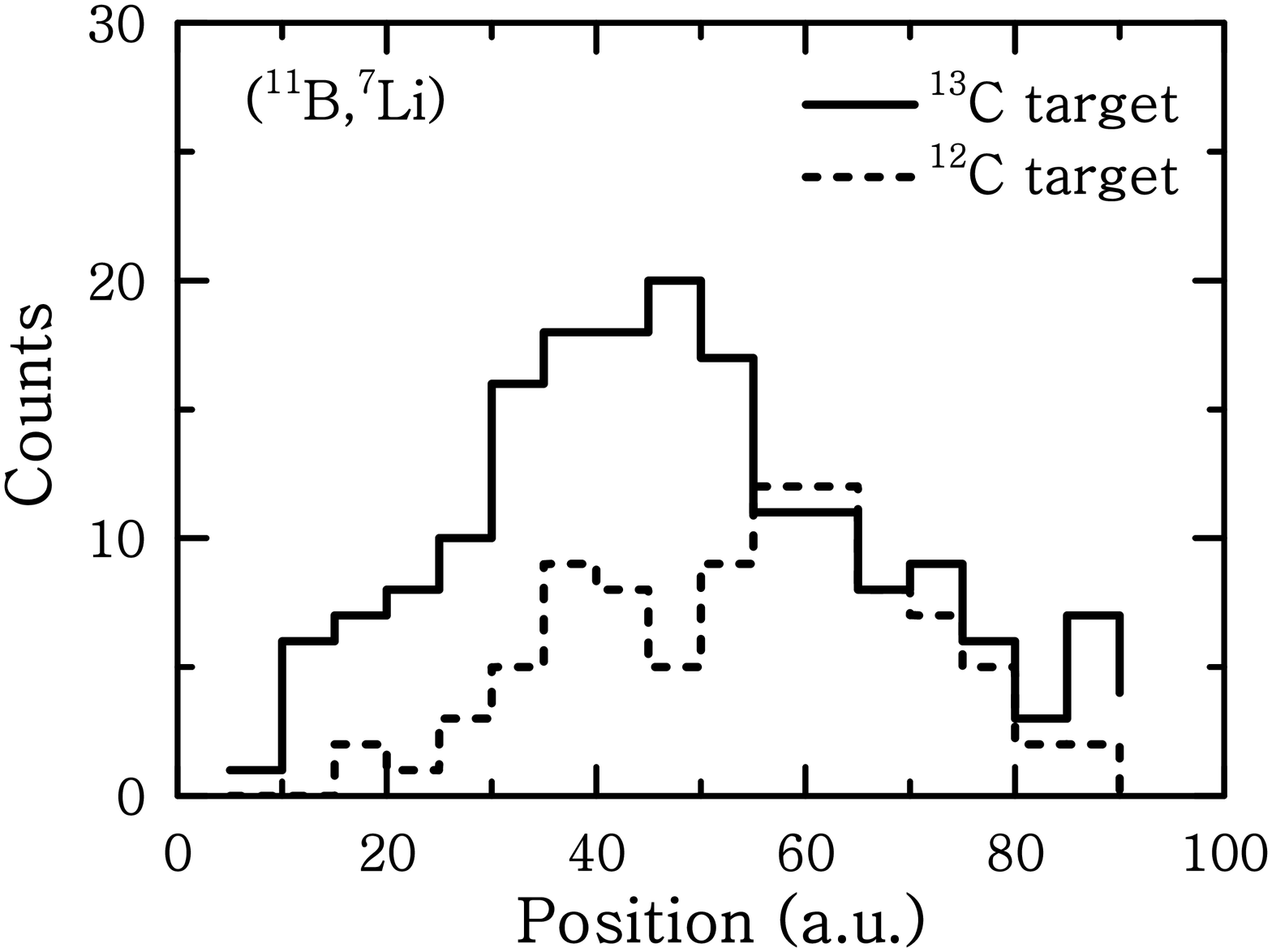} \caption{Focal-plane position
spectra of the $^{7}$Li events at $\theta_\mathrm{lab}$ = 10$^\circ$
from the $\alpha$-transfer reactions. The solid and dashed lines
denote the results from the enriched $^{13}$C target and natural
$^{12}$C target, respectively. \label{fig1}}
\end{figure}

\begin{figure}
\epsscale{.80} \plotone{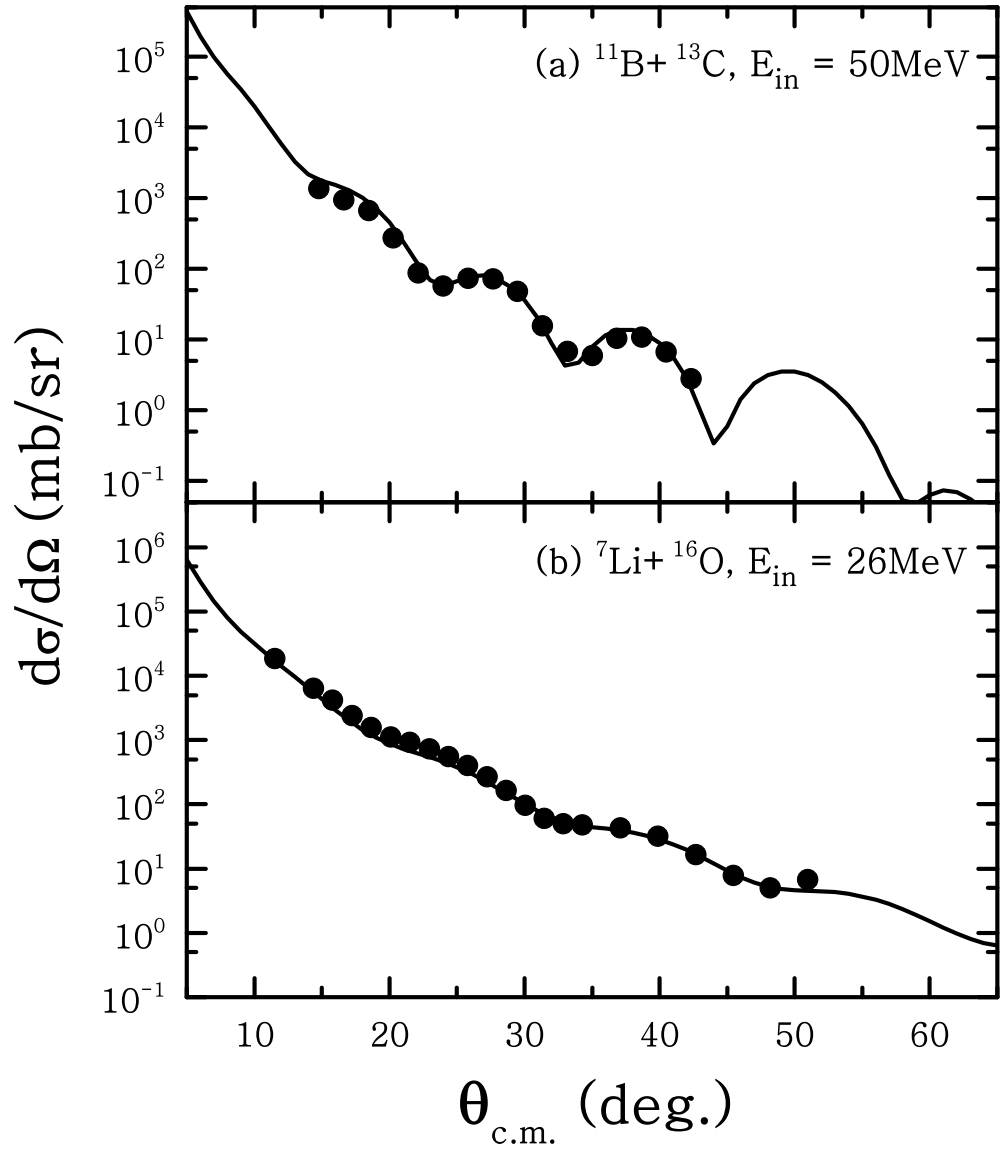} \caption{Angular distributions of
the $^{11}$B+$^{13}$C elastic scattering at incident energy of 50
MeV and the $^{7}$Li+$^{16}$O elastic scattering at incident energy
of 26 MeV. The solid curves represent the calculations with the
fitted OMP parameters. \label{fig2}}
\end{figure}

\begin{figure}
\epsscale{.90} \plotone{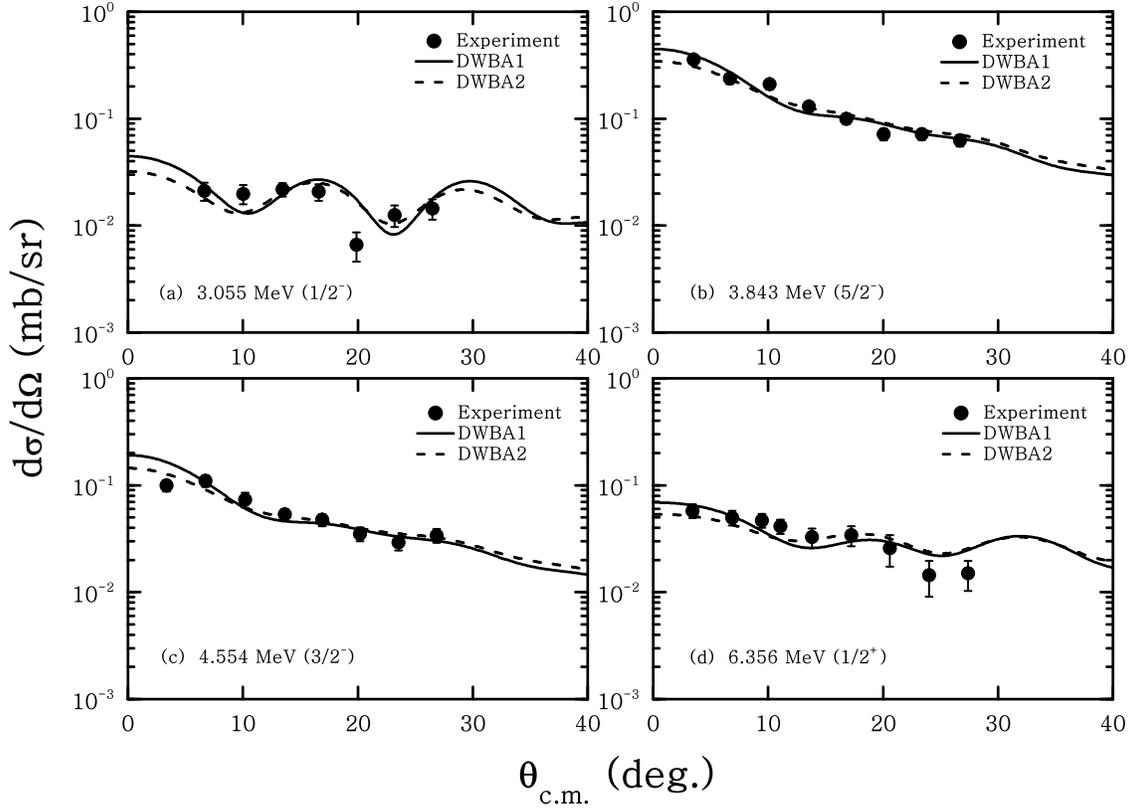} \caption{Angular distributions of
the $^{13}$C($^{11}$B,\,$^{7}$Li)$^{17}$O reaction leading to the
excited states at $E_x$ = 3.055 MeV, 3.843 MeV, 4.554 MeV and 6.356
MeV. The curves denote the DWBA calculations with the fitted OMP
parameters. DWBA1 and DWBA2 represent the results using the
spectroscopic amplitudes of $^{11}$B from \citet{rud05} and
\citet{kur73}, respectively. \label{fig3}}
\end{figure}

Due to the presence of $^{12}$C in the $^{13}$C target, the $^7$Li
events from the $^{13}$C($^{11}$B,\,$^{7}$Li)$^{17}$O$^*$ (6.356
MeV) reaction were mixed with those from the
$^{12}$C($^{11}$B,\,$^{7}$Li)$^{16}$O$^*$ (6.917 MeV) reaction. To
evaluate this background the ($^{11}$B,\,$^{7}$Li) reactions were
measured for both the $^{13}$C and $^{12}$C targets at each angle
with the same experimental setup. As an example, in Fig. \ref{fig1}
we display the focal-plane position spectra of $^7$Li at
$\theta_\mathrm{lab}$\,=\,10$^\circ$ from the ($^{11}$B,\,$^{7}$Li)
transfer reactions. The background from $^{12}$C accounts for
approximately half of total events from the $^{13}$C target. After
background subtraction and beam normalization, the angular
distributions of the elastic scattering and the
$^{13}$C($^{11}$B,\,$^{7}$Li)$^{17}$O$^*$ (6.356 MeV) reaction were
obtained, as presented in Figs. \ref{fig2} and \ref{fig3}. In Fig.
\ref{fig3} we also display the angular distributions of the
$^{13}$C($^{11}$B,\,$^{7}$Li)$^{17}$O reaction leading to the other
three states ($E_x$ = 3.055 MeV, 3.843 MeV, and 4.554 MeV). Note
that the measurements for these three states were not affected by
the $^{12}$C background.

\subsection{Extraction of the ANC}

The finite-range distorted wave Born approximation (DWBA) method
with the FRESCO code \citep{tho88} was used to analyze the
experimental angular distributions of the transfer reaction. The
optical model potential (OMP) parameters for the entrance and exit
channels were obtained by fitting the experimental angular
distributions of the $^{11}$B\,+\,$^{13}$C and $^{7}$Li\,+\,$^{16}$O
elastic scattering, respectively (Fig. \ref{fig2}). Full complex
remnant term interactions were included in the transfer reaction
calculations. The parameters of the core-core
($^{7}$Li\,+\,$^{13}$C) potential of \citet{coo87} gave reasonable
account of the elastic scattering of $^7$Li from $^{13}$C at 34 MeV.
The parameters used in the DWBA calculations are listed in Table
\ref{tab1}.

To obtain the spectroscopic factor ($S_\alpha$) and the ANC of
$\alpha$-cluster in $^{17}$O the spectroscopic amplitudes of
$\alpha$-cluster in the ground state of $^{11}$B need to be fixed.
The single-particle wave function describing the relative motion
between $\alpha$-cluster and $^7$Li core in the $^{11}$B ground
state can have two components denoted by quantum numbers $NL_j$ =
3$S_0$ and 2$D_2$, respectively, where $N$ is the number of radial
nodes that include the origin but not the infinity and $L$ and $j$
are the orbital and total angular momenta, respectively. The
spectroscopic amplitudes of these two components are $-$0.509 and
0.629, respectively, from a shell model calculation \citep{kur73},
and $-$0.638 and $-$0.422, respectively, from the translationally
invariant shell model \citep{rud05}. In the present analysis, both
sets of spectroscopic amplitudes were used and the resulting
difference was incorporated in the total uncertainty of our result.

The $\alpha$-cluster single-particle wave functions were calculated
using conventional Woods-Saxon potentials whose depths were adjusted
to reproduce the binding energies of $\alpha$-cluster in the ground
state of $^{11}$B and in the four states of $^{17}$O. The quantum
numbers $NL_j$ for $\alpha$-cluster in the 6.356 MeV 1/2$^+$ state
of $^{17}$O were fixed to be $4P_1$ by the oscillatory energy
conservation relation $2(N-1)+L = \sum_{i=1}^4 2(n_i-1)+l_i$, where
$n_i,l_i$ are the corresponding single-nucleon shell quantum
numbers. In addition, the quantum numbers for $\alpha$-cluster in
the other three states, 3.055 MeV 1/2$^-$, 3.843 MeV 5/2$^-$, and
4.554 MeV 3/2$^-$, were determined to be $4S_0$, $3D_2$, and $3D_2$,
respectively.

The geometry parameters, radius $r_0$ and diffuseness $a$, of the
Woods-Saxon potential for $\alpha$-cluster in $^{11}$B were adjusted
to give the root-mean-square (RMS) radius ($\sqrt{\langle
r^2\rangle}=3.204$ fm) of the $\alpha$-cluster wave function. This
was calculated using the following relation between cluster sizes
and their mean distances within a nucleus,
\begin{equation}
\langle r_{\textrm{B}}^{2} \rangle = {m_{\textrm{He}} \over
m_{\textrm{B}}} \langle r_{\textrm{He}}^{2} \rangle+{m_{\textrm{Li}}
\over m_{\textrm{B}}} \langle r_{\textrm{Li}}^{2}
\rangle+{m_{\textrm{He}}m_{\textrm{Li}} \over m_{\textrm{B}}^2}
\langle r^{2} \rangle, \label{eq1}
\end{equation}
where the RMS radii of $^4$He, $^7$Li, and $^{11}$B were taken to be
1.47 fm, 2.384 fm, and 2.605 fm, respectively \citep{lia90}. The
resulting parameters are $r_0\,=\,0.92$ fm and $a = 0.65$ fm. We
investigated the dependence of the calculated $S_\alpha$ and the ANC
on the geometry parameters for $^{11}$B. With a diffuseness between
0.65\,$-$\,0.75 fm, the radius was adjusted to reproduce the RMS
radius of 3.204 fm. The impact of this change on the $S_\alpha$ was
found to be less than 2\%. We also investigated the dependence of
the $S_\alpha$ and the ANC on the geometry parameters for $^{17}$O
using the same range of $r_0$ (0.9\,$-$\,1.06 fm) and $a$
(0.60\,$-$\,0.76 fm) that was selected by maximum likelihood
function set at 3$\sigma$ level by \citet{pel08}. Using steps of
0.02 fm for both $r_0$ and $a$, 81 sets of geometry parameters were
obtained and used to calculate 81 values of the $S_\alpha$ and the
ANC. Their standard deviations were taken as the uncertainty
deriving from the geometry parameters for $^{17}$O. The parameters
$r_0$\,=\,1.00 fm and $a$\,=\,0.76 fm provide the best description
for the angular distributions of the four states (see also Fig.
\ref{fig3}).

In Fig. \ref{fig3} we display the calculated angular distributions
normalized to the experimental data for the
$^{13}$C($^{11}$B,\,$^{7}$Li)$^{17}$O reaction populating the 3.055
MeV, 3.843 MeV, 4.554 MeV, and 6.356 MeV states in $^{17}$O. For the
3.055 MeV state the $S_\alpha$ factor was found to be
0.19\,$\pm$\,0.06, compatible with the values obtained from the
($^{6}$Li,\,$d$) reaction ($S_\alpha$\,$\approx$\,0.18\,$-$\,0.3) by
\citet{kee03} and from the ($^{7}$Li,\,$t$) reaction
($S_\alpha$\,=\,0.27\,$\pm$\,0.05) by \citet{pel08}. For the 3.843
MeV and 4.554 MeV states the $S_\alpha$ factors were found to be
0.078\,$\pm$\,0.025 and 0.060\,$\pm$\,0.019, respectively. These
disagree with the values of 0.19\,$-$\,0.34 and 0.27\,$-$\,0.48,
respectively for the two states, given by \citet{kee03}, who adopted
geometry parameters $r_0$\,=\,1.25 fm and $a$\,=\,0.65 fm for
$^{17}$O and used a coupled-channel calculation to fit the data.
When using these geometry parameters we were not able to reproduce
our measured $^{13}$C($^{11}$B,\,$^{7}$Li)$^{17}$O angular
distributions of the four states and obtained an extremely small
value of maximum likelihood function. For the 4.554 MeV state our
result is consistent with that of \citet{pel08}, who found an
$S_\alpha$ factor of 0.10\,$\pm$\,0.05 via the ($^{7}$Li,\,$t$)
reaction.

The $S_\alpha$ for the 6.356 MeV 1/2$^+$ state of $^{17}$O was
derived to be 0.37\,$\pm$\,0.12. The error results from the
statistics (23\%), the target thickness (8\%), the uncertainties
from the spectroscopic amplitudes (3\%) for $^{11}$B, the geometry
parameters (2\%) for $^{11}$B, and the geometry parameters (20\%)
for the 1/2$^+$ state of $^{17}$O. The square of the Coulomb
modified ANC ($\tilde{C}^2$) was then extracted to be 4.0 $\pm$ 1.1
fm$^{-1}$ using the relation, $\tilde{C}^2 = S_\alpha
{R^2\phi(R)^2/\tilde{W}(R)^2}$, where $\phi(R)$ is the radial
single-particle wave function for $\alpha$-cluster in the 1/2$^+$
state of $^{17}$O, and $\tilde{W}(R)=W(R)\Gamma(L+1+\eta)$ is the
Coulomb modified Whittaker function, with $\Gamma(L+1+\eta)$ being
gamma function and $\eta$ the Coulomb parameter. The uncertainty in
the ANC of 27.5\% is smaller than that in the $S_\alpha$ of 32.4\%.
This is because the variation of the geometry parameters yields a
change in $\phi(R)$ that is opposite in sign to the change in
$S_\alpha$ so the uncertainty from the $^{17}$O geometry parameters
in $\tilde{C}^2$ of 12\% is smaller than that in $S_\alpha$ of 20\%.

\subsection{Different evaluations of the ANC}

Three independent measurements in addition to the present work have
been performed to date to study the ANC or the $S_\alpha$ of the
1/2$^+$ state in $^{17}$O. A very small spectroscopic factor
($S_\alpha$\,$\sim$\,0.011) was found via measurement of the
$^{13}$C($^{6}$Li,\,$d$)$^{17}$O angular distribution with an
incident energy of 60 MeV \citep{kub03}. This indicated that the
contribution of the 1/2$^+$ subthreshold resonance is negligible.
However, a reanalysis of the same experimental data showed that the
DWBA analysis of \citet{kub03} could be flawed \citep{kee03}. These
authors derived larger $S_\alpha$ factors of 0.36 and 0.40 via DWBA
and coupled reaction channels (CRC) calculations, respectively.
\citet{joh06} measured the $^{6}$Li($^{13}$C,\,$d$)$^{17}$O angular
distribution at sub-Coulomb energies (8 and 8.5 MeV) and derived
$\tilde{C}^2$ = 0.89\,$\pm$\,0.23 fm$^{-1}$. \citet{pel08} measured
the $^{13}$C($^{7}$Li,\,$t$)$^{17}$O angular distribution with two
incident energies (28 and 34 MeV). The $S_\alpha$ and $\tilde{C}^2$
were found to be 0.29\,$\pm$\,0.11 and 4.5\,$\pm$\,2.2 fm$^{-1}$,
respectively.

In Table \ref{tab2} we list the $S_\alpha$ and $\tilde{C}^2$ values
from the present study and from the literature sources mentioned
above. Good agreement for the $S_\alpha$ and $\tilde{C}^2$ is found
between \citet{kee03}, \citet{pel08}, and the present work, which
used three different transfer systems and covered an energy range of
4\,$-$\,10 MeV/u. This indicates that the $S_\alpha$ and ANC for the
1/2$^+$ state of $^{17}$O should not significantly depend on the
selection of transfer systems and incident energies. On the other
hand, the $\tilde{C}^2$ obtained from the $^{6}$Li($^{13}$C,\,$d$)
reaction at sub-Coulomb energies by \citet{joh06} is about five
times smaller than that obtained in the present work and by
\citet{pel08}. The sub-Coulomb ($^{6}$Li,\,$d$) $\alpha$-transfer
cross section was recently remeasured to understand the source of
this discrepancy and the data analysis is in progress (G. V.
Rogachev 2012, private communication). It remains to be seen if this
analysis will result in a revised value of $\tilde{C}^2$.

\section{The $^{13}$C($\alpha$,\,$n$)$^{16}$O reaction rate}

The astrophysical S-factor of the $^{13}$C($\alpha$,\,$n$)$^{16}$O
reaction via the resonances can be calculated with the Breit-Wigner
formula,
\begin{eqnarray}
S(E)=\pi{\hbar^{2} \over 2\mu}\frac{2J_R + 1}{(2J_p + 1)(2J_t +
1)}\frac{\Gamma_{\alpha}(E)\Gamma_n(E+Q)}{(E-E_{R})^{2}+(\Gamma_{tot}/2)^{2}}\exp(2\pi\eta),\label{eq2}
\end{eqnarray}
where $\mu$ is the reduced mass of the $\alpha$\,+\,$^{13}$C system;
$E_{R}$ represents the resonance energy; $J_R$, $J_p$, and $J_t$ are
the spins of the excited states in $^{17}$O, $\alpha$, and $^{13}$C,
respectively; $\Gamma_{\alpha}$, $\Gamma_{n}$, and $\Gamma_{tot}$
denote the $\alpha$-, neutron-, and total widths, respectively; and $Q$
is the reaction $Q$-value of $^{13}$C($\alpha$,\,$n$)$^{16}$O.

For the resonances above the $\alpha$-threshold, the energy
dependence of these partial widths is given by
\begin{eqnarray}
\Gamma_{\alpha}(E)=\Gamma_{\alpha}(E_{R}){P_{l_i}(E) \over
P_{l_i}(E_R)},\label{eq3}
\end{eqnarray}
and
\begin{eqnarray}
\Gamma_{n}(E+Q)=\Gamma_{n}(E_{R})\Big{(}{E+Q \over
E_{R}+Q}\Big{)}^{l_f+1/2},\label{eq4}
\end{eqnarray}
where $\Gamma_{\alpha}(E_{R})$ and $\Gamma_{n}(E_{R})$ denote the
experimental $\alpha$- and neutron-widths, respectively; $P_{l_i}(E)$ is the
$\alpha$-penetrability; and $l_i$ and $l_f$ are the orbital angular
momenta for $\alpha$ and neutron in the excited states of $^{17}$O,
respectively \citep[see, e.g.][]{ili07}.

For the subthreshold resonance, the energy dependence of the
neutron-width can be also obtained by Eq. \ref{eq4}, while the
dependence of the $\alpha$-width is expressed as
\begin{eqnarray}
\Gamma_{\alpha}(E)=2\gamma_\alpha^2P_{l_i}(E). \label{eq5}
\end{eqnarray}
Here the reduced $\alpha$-width $\gamma_\alpha^2$ can be given by
\begin{eqnarray}
\gamma_\alpha^2 = {\hbar^2R_c \over 2\mu} S_\alpha
\phi(R_c)^2={\hbar^2 \over 2\mu R_c} \tilde{C}^2 \tilde{W}(R_c)^2,
\label{eq6}
\end{eqnarray}
and was extracted to be 12.7\,$\pm$\,3.5 keV at the channel
radius $R_c$\,=\,7.5 fm. This large radius was chosen to reach the
Coulomb asymptotic behavior of $\phi(R)$, as suggested by
\citet{des87} and \citet{pel08}.

In Table \ref{tab3} we list the resonant parameters for the 1/2$^+$
subthreshold state employed to obtain the astrophysical S-factor of
the 1/2$^+$ subthreshold resonance in
$^{13}$C($\alpha$,\,$n$)$^{16}$O. The uncertainties in the S-factor
were investigated by varying $\gamma_\alpha^2$, $E_{R}$, and
$\Gamma_{n}(E_{R})$. We also investigated the dependence of the
S-factor on the channel radius by changing $R_c$ from 6 fm to 7.5
fm. The S-factor at the Gamow peak of 190 keV, $S(190)$, was derived
to be (8.4\,$\pm$\,2.3)$\times$10$^5$ MeV b. The error is dominated
by the uncertainty in the resonant parameters for the 1/2$^+$
subthreshold state. The uncertainty from the channel radius is 4\%
in $S(190)$, varying slightly with the energy.

We calculated the total S-factor of $^{13}$C($\alpha$,\,$n$)$^{16}$O
by including the properties of the $^{17}$O states up to 8.342 MeV
from the compilation of \citet{til93}, and considering their
interferences. However, the resulting total S-factor did not agree
with the data from the direct measurements of \citet{dro93}, who
measured the S-factor at the lowest energies to date, and of
\citet{hei08}, who performed a systematical experimental
verification of neutron efficiency over a range of well-defined
energies. Moreover, the resonant structure near the 3/2$^+$
resonance at $E_R$\,=\,0.842 MeV was not well reproduced using the
parameters from \citet{til93}. Hence, we adjusted the $\alpha$- and
neutron-widths for this state to provide the best fitting of the
experimental data. The fitting resulted in broader partial widths
[$\Gamma_{\alpha}(E_{R})$\,=\,0.08 keV, $\Gamma_{n}(E_{R})$\,=\,340
keV] than the recommended values of \citet{til93}
[$\Gamma_{\alpha}(E_{R})$\,=\,0.07 keV, $\Gamma_{n}(E_{R})$\,=\,280
keV], in agreement with \citet{pel08} and \citet{hei08}. All
resonant parameters used in the present calculations are listed in
Table \ref{tab3}.

\begin{figure}
\epsscale{.80} \plotone{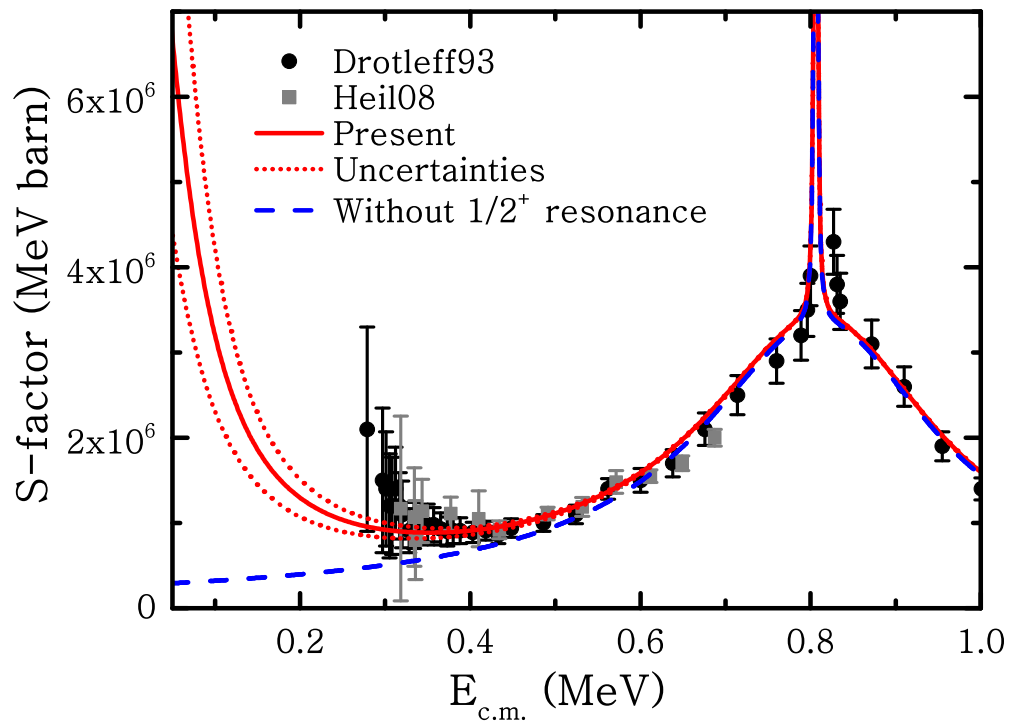} \caption{Astrophysical S-factor
for the $^{13}$C($\alpha$,\,$n$)$^{16}$O reaction. Dots and squares
represent the data of \citet{dro93} and \citet{hei08}, respectively.
The solid and dotted (red) curves represent the present S-factor and
its uncertainties. The dashed (blue) curve represents the S-factor
by excluding the contribution of the 1/2$^+$ subthreshold resonance.
(A color version of this figure is available in the online journal.)
\label{fig4}}
\end{figure}

In Fig. \ref{fig4} we display the resulting astrophysical
$^{13}$C($\alpha$,\,$n$)$^{16}$O S-factor as a function of the energy in
the center of mass frame ($E_\mathrm{c.m.}$). The uncertainty mainly
results from the resonant parameters of the 1/2$^+$ subthreshold state
since the resonant parameters of the states above the $\alpha$-emission
threshold are relatively well constrained by fitting the experimental
data. The contribution of the 1/2$^+$ resonance results in a clear
increase of the S-factor at lower energies. At the Gamow peak of 190
keV the 1/2$^+$ subthreshold resonance dominates the
$^{13}$C($\alpha$,\,$n$)$^{16}$O reaction. At this energy the
contribution of this resonance accounts for 61\% of total S-factor.

The astrophysical $^{13}$C($\alpha$,\,$n$)$^{16}$O reaction rate
was calculated with the present S-factor using:
\begin{eqnarray}
\label{eq7}%
N_A \langle\sigma v\rangle = N_A\Big{(}{8 \over
\pi\mu}\Big{)}^{1/2}{1 \over (kT)^{3/2}}\int^{\infty}_0
S(E)\exp\Big{[}-{E \over kT}-2\pi\eta\Big{]}dE,
\end{eqnarray}
where $N_{A}$ is Avogadro's number.
In Table \ref{tab4} we list the present adopted
rate as function of the temperature together with its
upper and lower limits. In Fig. \ref{fig5} we compare the
present rate with the previous compilations by CF88 \citep{cau88}
and NACRE \citep{ang99} and with other rates available in the
literature \citep{den95,kub03,joh06,pel08,hei08} at the temperature of 100 MK.
Our rate agrees with that of \citet{den95} and it
is about two times smaller than the NACRE recommended rate, but within
its lower limit. Both the rates by NACRE and \citet{den95}
were derived from an
extrapolation of the lowest energy experimental data available to date of
\citet{dro93}. There is a large discrepancy of up to a
factor of 2 between the present rate and the rates by CF88, \citet{kub03},
and \citet{joh06} due to the fact that CF88 did not take into account
the contribution of the
1/2$^+$ subthreshold resonance and \citet{kub03} and \citet{joh06} found
a significantly smaller contribution than ours of this resonance to the rate.
Our rate is in good agreement
with that of \citet{pel08}, who found the contribution of the
subthreshold resonance similar to ours, and is consistent with that of
\citet{hei08}, who used an extensive multichannel $R$-matrix analysis
to constrain the possible contribution from subthreshold
resonances by taking into account all open reaction channels for the
$^{13}$C\,+\,$\alpha$ and $^{16}$O\,+\,$n$ systems.

For conveniency, we fitted our rate with the expression used in the
astrophysical reaction rate library REACLIB \citep{thi87,rau01},
\begin{eqnarray}
\label{eq8}%
N_A \langle\sigma v\rangle =
\exp[a_1+a_2T_{9}^{-1}+a_3T_{9}^{-1/3}+a_4T_{9}^{1/3}
+a_5T_{9}+a_6T_{9}^{5/3}+a_7\ln(T_{9})]\nonumber\\
+\exp[a_8+a_9T_{9}^{-1}+a_{10}T_{9}^{-1/3}+a_{11}T_{9}^{1/3}
+a_{12}T_{9}+a_{13}T_{9}^{5/3}+a_{14}\ln(T_{9})].
\end{eqnarray}
Here $T_9$ is the temperature in units of 1 GK. The coefficients
$a_i$ for our adopted value and lower and upper limits of the
$^{13}$C($\alpha$,\,$n$)$^{16}$O reaction rate are listed in Table
\ref{tab4}. The overall fitting errors are less than 7\% at
temperatures from 0.04 to 10 GK.

\section{Astrophysical implications for $s$-process nucleosynthesis in AGB
stars}

To assess and understand the impact of the new rates of the
$^{13}$C($\alpha$,\,$n$)$^{16}$O reaction we used a post-processing
code where a nuclear network of 320 species (from H to Bi) and 2,336
reactions is solved simultaneously with mixing within the star when
convective regions are present. We used stellar structure inputs,
such as temperature, density, and convective velocity, calculated
previously by the Stromlo stellar structure code \citep{lat86}
including mass-loss on the AGB phase with the prescription of
\citet{vas93}. We included the formation of the $^{13}$C pocket by
artificially allowing an exponentially decreasing proton profile to
form just below the base of the convective envelope at the end of
each TDU episode over a mass of 0.002 $M_\sun$ (roughly 1/10$^{\rm
th}$ of the mass of the He intershell, in the low-mass AGB models
considered here). The details of this procedure and the codes used
to compute the models have been described previously
\citep[e.g.,][]{lug04,kar09}.

We considered four models, which are summarized in Table \ref{tab5}:
a $M\,=\,$3 $M_\sun$ model of metallicity $Z\,=\,$0.02 similar to
that discussed by \citet{lug03a}, a $M\,=\,$1.8 $M_\sun$ model of
$Z\,=\,$0.01 from \citet{kar10}, and $M\,=\,$1 $M_\sun$ and
$M\,=\,$1.5 $M_\sun$ models of $Z\,=\,$0.0001 from \citet{lug12}. We
selected these models for three reasons. (1) Models of higher masses
probably do not experience the $^{13}$C neutron source as protons
burn while being ingested in the He intershell, preventing the
formation of the $^{13}$C pocket \citep{gor04,her04}. It is still
not clear what the effect of this burning is on the whole stellar
structure and the $s$-process, but it seems reasonable to us for the
time being to not include a $^{13}$C pocket for masses larger than
$\sim$\,4 $M_\sun$. In these massive AGB stars, the temperature in
the TPs exceeds 300 MK so that the
$^{22}$Ne($\alpha$,\,$n$)$^{25}$Mg reaction is activated and likely
plays the role of the main neutron source \citep{van12}. (2) The
$^{13}$C in the pocket normally burns before the onset of the
following TP \citep{str95}, however, in some cases the temperature
may not be high enough for this to happen and some $^{13}$C could be
left over to burn in the following TP \citep[see also][]{cri09a}. In
our lowest-mass models the first few $^{13}$C pockets are ingested
in the following TP while they still contain a relatively large
amount of $^{13}$C. These models are thus qualitatively different
from the $M\,=\,$3 $M_\sun$ model where $^{13}$C always burns before
the onset of the next TP. In details, we have 6, 15, and 2 TDU
episodes in the $M\,=\,$1.8 $M_\sun$ model of $Z\,=\,$0.01, and the
$M\,=\,$1.5 $M_\sun$ and $M\,=\,$1 $M_\sun$ models of
$Z\,=\,$0.0001, respectively. These numbers represent also the
number of $^{13}$C pockets introduced in each models, since a
$^{13}$C pocket is introduced after each TDU episode. Of these
$^{13}$C pockets, the first 4 and 3 for the $M\,=\,$1.8 $M_\sun$ and
$M\,=\,$1.5 $M_\sun$ models, respectively, are ingested in the
following TP while the abundance of $^{13}$C by number is still
higher than 10$^{-4}$. The same occurs for both the $^{13}$C pockets
introduced in the $M\,=\,$1 $M_\sun$ model. As the AGB star evolves
the temperature in the He intershell increases so that this effect
disappears for later $^{13}$C pockets. The exact number of ingested
$^{13}$C pockets relative to their total number increases with
decreasing the stellar mass and increasing the metallicity. This is
because the temperature decreases for lower masses and higher
metallicities and thus the chance of some unburnt $^{13}$C surviving
to be ingested in the next TP is higher. (3) Finally the low-mass
and low-metallicity models (1 $M_\sun$ and 1.5 $M_\sun$,
$Z$\,=\,0.0001) present a further regime for the activation of the
$^{13}$C neutron source. Proton ingestion episodes occur in the
first few TPs during which a small amount of protons are ingested
directly inside the TPs and thus some extra $^{13}$C is produced and
burnt inside these convective regions \citep{cri09b,lug12}. In Fig.
\ref{fig6} we schematically illustrate the location in time and
space where the different regimes operate within the
thermally-pulsating structure of an AGB star and indicate the models
where the regimes occur.

In Fig. \ref{fig7} we display the variations of the abundance of
some selected $s$-process elements and isotopic ratios at stellar
surface at the end of the evolution obtained from our models using
the present $^{13}$C($\alpha$,\,$n$)$^{16}$O rate with respect to
the two compilations, CF88 and NACRE, respectively (see Fig.
\ref{fig5} for a comparison of reaction rates). The results obtained
using the rates from \citet{pel08}, \citet{hei08}, and \citet{den95}
agree with the present rate within 10\% for all the models. The rate
from \citet{joh06} produces larger variations, but always within
those reported for the CF88 and NACRE rates. Using the rate from
\citet{kub03} on the other hand produces variations outside of the
ranges in Fig. \ref{fig7}, e.g., Pb is up to 40\% lower than the
value obtained using the present rate. We chose to plot the elements
representing the first $s$-process peak (Sr), the second $s$-process
peak (Ba), and the third $s$-process peak (Pb). These can be
observed in stars and provide a description of the overall
$s$-process distribution, which is a function of the total
time-integrated neutron flux. We further plot two isotopic ratios:
$^{96}$Zr/$^{94}$Zr and $^{86}$Kr/$^{82}$Kr. These are sensitive to
the activation of the branching points at $^{95}$Zr and $^{85}$Kr,
respectively, and thus to the local value of the neutron density.
These ratios can be measured in meteoritic stardust silicon carbide
(SiC) grains that originated in C-rich AGB stars. We also report on
the production of fluorine, however, we find that its production is
not significantly sensitive (by $\simeq$10\% at most) to the choice
of the $^{13}$C($\alpha$,\,$n$)$^{16}$O reaction in all four models
\citep[see also][]{lug04}.

In the 1.8 $M_\sun$, $Z\,=\,$0.01 model we find that the overall
$s$-process distribution changes when different
$^{13}$C($\alpha$,\,$n$)$^{16}$O rates are used. The present rate
produces 16\% more Ba and 25\% more Pb than the slower CF88 rate. On
the other hand, the present rate results in a 14\% lower abundance
of Pb than the faster NACRE rate. These differences are a signature
of a different efficiency of the $s$-process. A lower efficiency
produces less Ba and Pb than Sr because for the neutron capture flux
to reach Ba and Pb higher number of neutrons per Fe seed are
required. Slower rate results in more $^{13}$C left over in the
pocket to be engulfed in the following TP and the $s$-process is
less efficient when $^{13}$C is engulfed in the TP than when
$^{13}$C burns radiatively for two reasons. (i) In the radiative
$^{13}$C pocket there is no mixing among the different layers of the
pocket and thus $^{14}$N nuclei produced in the region of the pocket
where the initial number of proton is higher than 0.01 are not mixed
to the $^{13}$C-rich layer. This produces the highest $s$-process
efficiency because $^{14}$N is a neutron poison via the
$^{14}$N($n,\,p$)$^{14}$C reaction, which removes neutrons from
being captured by Fe seeds and their progeny. In the convective TP
instead $^{13}$C nuclei are mixed with the $^{14}$N present in the
pocket as well as in the H burning ashes. (ii) The neutrons in the
radiative $^{13}$C pocket are released over a very small mass (0.002
$M_\sun$ in our models), there are more neutrons present locally per
initial Fe seed and the flux of neutron captures can reach to the
heaviest elements up to Pb. In the convective TP instead the
neutrons are released over a larger mass ($\sim$\,0.01$-$0.02
$M_\sun$ in our models), there are less neutrons per Fe seed
resulting in the production of the lighter $s$-process elements like
Sr. The isotopic ratios plotted in Fig. \ref{fig7} behave in the
opposite way of Pb because the local neutron density is higher, and
thus branching points are more activated, when $^{13}$C is ingested
in the TPs due to the higher temperature resulting in a shorter
burning timescale.

Interestingly, the low-metallicity $Z\,=\,0.0001$ models present the
opposite result to the $Z\,=\,0.01$ model. E.g., in the 1 $M_\sun$
model the present rate produces 25\% less Pb than the slower CF88
rate. These stellar models have lower masses and a much larger
fraction of $^{13}$C burns while ingested in the TP, together with
$^{14}$N, due to both incomplete burning of the $^{13}$C pocket
before the next TP and the proton ingestion episodes. As mentioned
above, $^{14}$N acts a neutron poison via the ($n,\,p$) channel, but
at the same time it is destroyed by $\alpha$-captures in the TP.
Hence, the timescale of $^{13}$C ingestion and burning with respect
to that of $^{14}$N has an impact on the final neutron flux. This
could explain why slower $^{13}$C($\alpha$,\,$n$)$^{16}$O rates
produce more free neutrons since if neutrons are released at later
times there is less $^{14}$N to capture them. In these conditions,
one should also consider the effect of recycling of the protons made
by the $^{14}$N($n,\,p$)$^{14}$C reaction. These protons can be
captured by the abundant $^{12}$C to make more $^{13}$C, but also by
$^{13}$C itself. It is difficult to evaluate analytically the final
outcome of all these combined effects. Overall, our numerical
models, which solve simultaneously mixing and burning in the TPs,
indicate that the present rate results in a lower overall neutron
flux than the CF88 rate. This is accompanied by a higher or a lower
neutron density, depending on the stellar mass and the exact ratio
of $^{13}$C burning convectively to $^{13}$C burning radiatively.
17\% lower and 22\% higher $^{86}$Kr/$^{82}$Kr ratios, which are
extremely sensitive to even small changes in the neutron density,
are obtained for the 1 $M_\sun$ and 1.5 $M_\sun$ models,
respectively.

In the 3 $M_\sun$, $Z\,=\,$0.02 model we do not find any variations
in the resulting abundances within 5\%. This is because in this
model $^{13}$C always completely burns before the onset of the next
TP. The total integrated neutron flux is thus determined only by the
initial amount of $^{13}$C in the pocket and not by how fast it
burns. We expect the same behavior for more massive AGB model if a
$^{13}$C pocket was to be considered in these cases. The isotopic
ratios in this model are also not sensitive to the
$^{13}$C($\alpha$,\,$n$)$^{16}$O rate as they depend mainly on the
neutrons released in the TPs when the $^{22}$Ne neutron source is
marginally activated.

It is interesting to compare the predicted abundances of the
$s$-process elements and isotopic ratios with the observations. We
find that some differences up to 25\% are predicted by the
$s$-process models, particularly for Pb, when varying the $^{13}$C
burning rates. Pb can be measured in carbon-enhanced metal-poor
(CEMP) stars in the Milky Way halo, which are believed to carry the
chemical signature of mass transfer from a more massive companion
during its AGB phase. However, the spectroscopic abundances in
$s$-process-enhanced stars are currently determined with typical
uncertainties around 80\% \citep[see Table 2 of][]{mas10}, larger
than the differences found here. In addition, the formation of the
$^{13}$C pocket itself is very uncertain and there is no agreement
on exactly which mechanism drives it, and on the impact of processes
such as rotation and magnetic fields. Also the details of the proton
ingestion episodes found in our low-metallicity models depend on the
treatment of convective borders in stars, one of the largest
uncertainties in stellar modeling. Future observations with high
resolution and high signal to noise ratio spectra of low-metallicity
stars, and further understanding of the formation of the $^{13}$C
pocket and the proton ingestion episodes will be all important for
better constraining the $s$-process model predictions.

Isotopic ratios affected by branching points are indicative of the
$^{13}$C burning rate, even though also in this case, other
uncertainties including the neutron-capture cross section of the
unstable nucleus at the branching point may have a more important
effect. The $^{96}$Zr/$^{94}$Zr ratio has been measured in S-type
AGB stars and in meteoritic stardust SiC grains that originated from
AGB stars of roughly solar metallicity. The data of
$^{96}$Zr/$^{94}$Zr from meteoritic grains are more precise
\citep[with uncertainties as low as 5\%$-$10\%,][]{lug03b} than the
data from S-type stars, where often it was only possible to derive
upper limits \citep{lam95}. The differences of 22\% found here when
changing the $^{13}$C($\alpha$,\,$n$)$^{16}$O rate do not
drastically change the interpretation of the data, which mostly
indicate strong deficits in $^{96}$Zr, with respect to solar, and
are overall matched by AGB models \citep{lug03b}. In the case of
$^{86}$Kr/$^{82}$Kr, stardust SiC data range from $\sim$0.5 to 2 of
the solar ratio of 1.52 \citep{lew94}. Our AGB models of metallicity
around solar ($Z$\,=\,0.01,\,0.02), which are believed to well
represent the parent stars of this stardust, only produce
$^{86}$Kr/$^{82}$Kr ratios lower than solar. However, \citet{ver04}
showed that the $^{86}$Kr atoms are probably implanted in SiC by the
high-velocity winds experienced during the evolution that follows
the AGB, the post-AGB and the planetary nebula nucleus phases. The
effect of the $^{13}$C($\alpha$,\,$n$)$^{16}$O reaction rate needs
to be evaluated with regards to the composition of Kr inside the He
intershell, which is exposed to the surface during these final
phases of the evolution instead of the surface abundances as done in
the present work. In this context the possibility of a proton
ingestion in the very late TP sometimes occurring during the
post-AGB evolution also has to be taken into account. First
simulations of this proton ingestion event by \citet{her11} have
shown that the $^{13}$C($\alpha$,\,$n$)$^{16}$O reaction rate plays
a main role in determining the final abundances (see their Fig. 12).
These models need to be further investigated in relation to the
present $^{13}$C($\alpha$,\,$n$)$^{16}$O reaction rate and the
$^{86}$Kr/$^{82}$Kr ratios in SiC grains.

\section{Summary and conclusion}

\begin{figure}
\epsscale{.70} \plotone{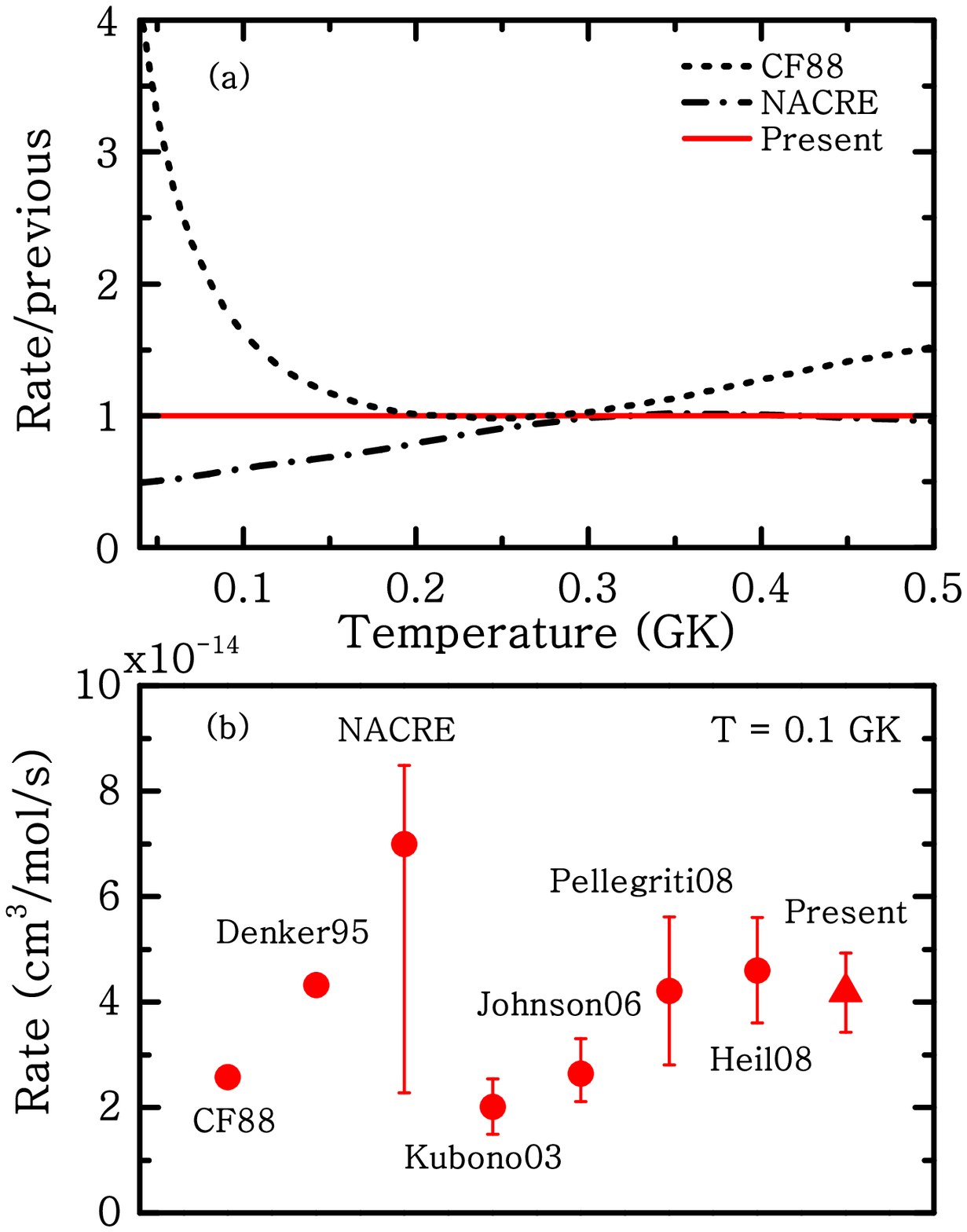} \caption{Comparison of the present
$^{13}$C($\alpha$,\,$n$)$^{16}$O rate with the previous results
available in the literature
\citep{cau88,den95,ang99,kub03,joh06,pel08,hei08}. (a) Ratio of the
present rate to the CF88 and NACRE compilations, in a temperature
range of 0.04\,$-$\,0.5 GK. (b) Comparison of the present rate with
those listed above at a temperature of 0.1 GK. (A color version of
this figure is available in the online journal.) \label{fig5}}
\end{figure}

\begin{figure}
\epsscale{.70} \plotone{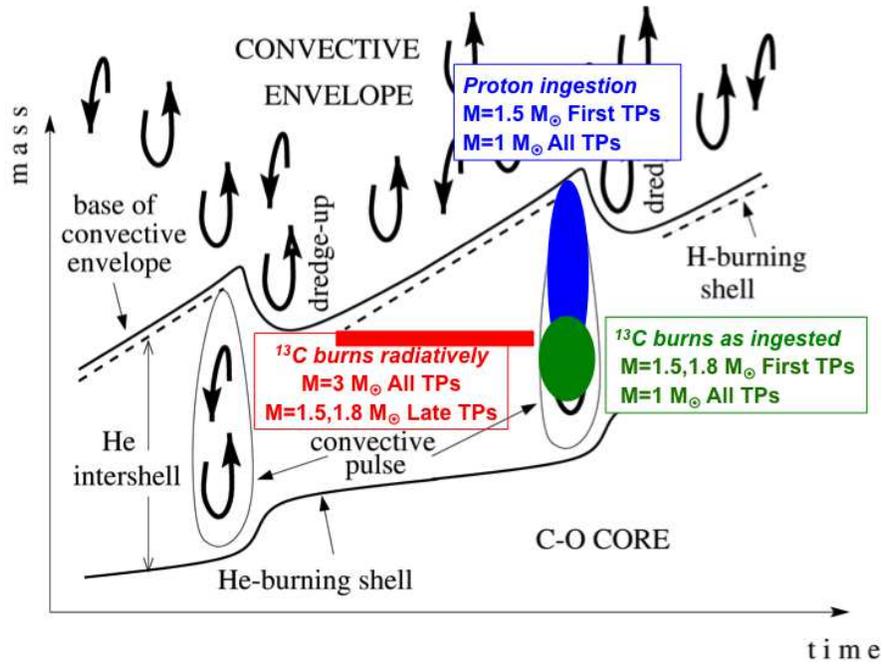} \caption{Schematic diagram of the
evolution of the structure of an AGB star with the superimposed
locations where neutrons are released by the $^{13}$C neutron source
in the different regimes described in the text. Also indicated are
the masses of the models and the typical TPs where each regime
occurs. (A color version of this figure is available in the online
journal.) \label{fig6}}
\end{figure}

\begin{figure}[htbp]
\epsscale{1.00} \plotone{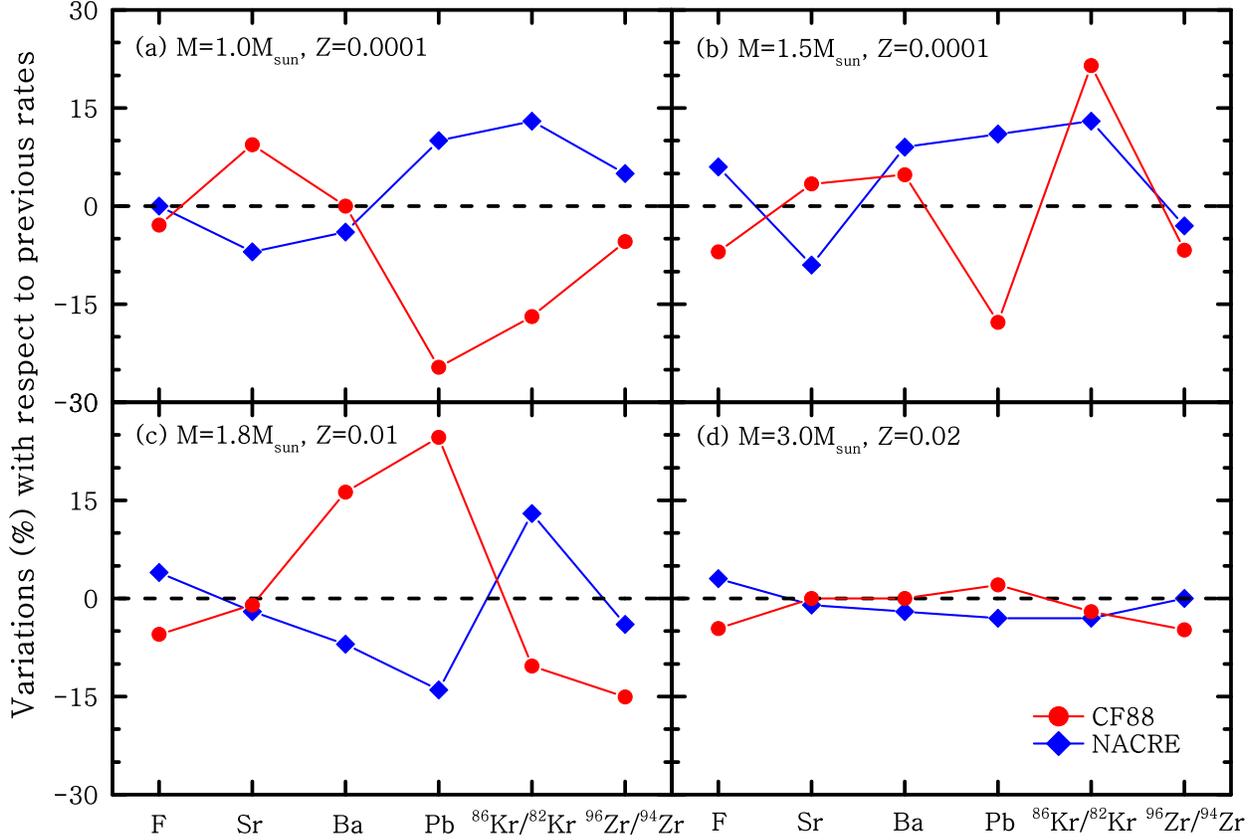} \caption{Percent of the
variations of the abundance and isotopic ratios when using the
present rate with respect to the CF88 and NACRE compilations. The
horizontal coordinates denote fluorine and the selected $s$-process
elements (Sr, Ba, Pb) and isotopic ratios ($^{86}$Kr/$^{82}$Kr,
$^{96}$Zr/$^{94}$Zr). Panels (a)-(d) show the results for the models
of $M=$1.0 $M_\sun$, $M=$1.5 $M_\sun$, $M=$1.8 $M_\sun$, and $M=$3.0
$M_\sun$, respectively. See text for details. (A color version of
this figure is available in the online journal.) \label{fig7}}
\end{figure}

We determined the stellar rate of the
$^{13}$C($\alpha$,\,$n$)$^{16}$O reaction and incorporated the new
reaction rate in calculations of the $s$-process nucleosynthesis in
AGB stars. The $S_\alpha$ and ANC for the 6.356 MeV 1/2$^+$
subthreshold state in $^{17}$O were obtained from the measurement of
the $^{13}$C($^{11}$B,\,$^{7}$Li)$^{17}$O angular distribution. This
provided an independent examination to shed some light on the
existing discrepancies in the $S_\alpha$ and ANC values derived from
different authors. Based on the measured ANC, we extracted the
$\alpha$-width of the 1/2$^+$ state in $^{17}$O, which is currently
the most uncertain parameter for determining the
$^{13}$C($\alpha$,\,$n$)$^{16}$O reaction rate. By using the present
$\alpha$-width and considering the properties of $^{17}$O states up
to 8.342 MeV as well as their interferences, we derived the
astrophysical S-factor and the stellar rate of the
$^{13}$C($\alpha$,\,$n$)$^{16}$O reaction. At a temperature of 100
MK the new rate is roughly two times larger than the CF88 value and
two times smaller than that recommended by NACRE (see Fig.
\ref{fig5}). Verification of the present result using other
independent techniques is desirable, e.g., the Trojan horse approach
\citep{spi99,muk08}, and the isospin symmetry approach based on a
measurement of the 1/2$^+$ 6.560 MeV state in $^{17}$F
\citep{tim07}. In addition, an extension of the experimental data of
$^{13}$C($\alpha$,\,$n$)$^{16}$O toward lower energies is highly
desirable, which can probably only be performed in an underground
laboratory, e.g., LUNA \citep[see][]{cos09}.

We incorporated different $^{13}$C($\alpha$,\,$n$)$^{16}$O
reaction rates in calculations of the $s$-process nucleosynthesis in
AGB stars and found that: (1) If $^{13}$C burns completely in
radiative conditions during the interpulse phase (as for stars of
initial mass greater than $\sim$2 $M_\sun$), there is no change in
the final results. (2) If some $^{13}$C burns instead inside the
convective TPs (for stars of initial mass lower than $\sim$2
$M_\sun$), we find changes of up to 25\% in the $s$-process results,
particularly for Pb. There are model uncertainties related to
the result of point (2): a)
when $^{13}$C burning in the TPs is due to incomplete burning of the
$^{13}$C during the interpulse period, the exact stellar mass and
metallicity range where incomplete burning of the $^{13}$C during
the interpulse period occurs, as well as TP numbers, and the exact
amount of $^{13}$C ingested, all depend sensitively on the
temperature and density in the $^{13}$C pocket, on the interpulse
period, and on the details of the inclusion of the $^{13}$C pocket.
b) When $^{13}$C burning in the TPs is due to ingestion of protons
directly inside the TP \citep[as in low-mass and low-metallicity
stars, as well as in post-AGB stars experiencing a late TP,
see][]{her11}, the amount of $^{13}$C and neutrons produced strongly
depends on the physical and numerical treatment of the mixing scheme
adopted, which is at present uncertain.

Due to these model uncertainties together with the fact that the
stellar observations have relatively large error bars, it is
presently not possible to conclude if the new rate of the
$^{13}$C($\alpha$,\,$n$)$^{16}$O reaction provides the best match to
the available observational constraints. This may be however
possible in the future, when development of recent 3D hydrodynamical
models of the proton ingestion episodes \citep{sta11} will allow a
better understanding of neutron production and the $s$-process
inside TPs to be compared to the composition of stellar observations
and stardust grains.

\acknowledgments

B.G. thanks Natasha Timofeyuk for helpful discussions on DWBA
calculation and isospin symmetry in mirror $\alpha$-decays, and
thanks Grigory Rogachev for useful discussions on the source of the
discrepancy between their results, and thanks Grigory Rogachev and
Fa\"{i}rouz Hammache for providing him with their reaction rates in
tabular form. M.L. acknowledges the support of the ARC via a Future
fellowship and of Monash University via a Monash Fellowship. We
thank the anonymous referee for many helpful comments and
suggestions. We acknowledge the staff of Tandem Accelerator for the
smooth operation of the machine. This work was supported by the
National Natural Science Foundation of China under Grant Nos.
11021504, 10735100 and 11035001, the National Basic Research Program
of China (New physics and technology at the limits of nuclear
stability), the Outstanding tutors for doctoral dissertations of
S\&T project in Beijing under Grant No.
YB20088280101.\\

\clearpage

\clearpage

\begin{table}
\begin{center}
\caption{OMP parameters used in the present DWBA calculation.
$E_{\textrm{in}}$ denotes the incident energy in MeV for the
relevant channels, $V$ and $W$ are the depths (in MeV) of the real
and imaginary parts of Woods-Saxon potential, and $r$ and $a$ are
the radius and diffuseness (in fm) of Woods-Saxon potential.
\label{tab1}}
\begin{tabular}{cccccccc}
\tableline\tableline
Channel & $E_{\textrm{in}}$ & $V$ & $r_{v}$ &
$a_{v}$ & $W$ &
$r_{w}$ & $a_{w}$\\
\tableline
$^{11}$B+$^{13}$C & 50.0 & 182.64 & 0.788 & 0.740 & 8.193 & 1.250 & 0.740 \\
$^{7}$Li+$^{17}$O & 26.0 & 114.20 & 0.737 & 0.719 & 34.602 & 0.997 & 0.764 \\
$^{7}$Li+$^{13}$C & 31.8 & 159.00 & 0.630 & 0.810 & 8.160 & 1.330 & 0.780 \\
\tableline
\end{tabular}
\end{center}
\end{table}

\begin{table}
\begin{center}
\caption{The $S_\alpha$ and ANC ($\tilde{C}^2$) as derived from
different measurements for the
1/2$^+$ subthreshold state of $^{17}$O. \label{tab2}}
\begin{tabular}{ccccc}
\tableline\tableline
Reference & Transfer system & Incident energy&$S_\alpha$ &$\tilde{C}^2$ \\
&& (MeV)&& (fm$^{-1}$)\\ \tableline
\citet{kub03} & $^{13}$C($^{6}$Li,$d$)&60&0.011&\\
\citet{kee03} & $^{13}$C($^{6}$Li,$d$)\tablenotemark{a}&60&0.36$-$0.40\tablenotemark{b}&\\
\citet{joh06} & $^{6}$Li($^{13}$C,$d$)&8.0 and 8.5&&0.89 $\pm$ 0.23\\
\citet{pel08} & $^{13}$C($^{7}$Li,$t$)&28 and 34&0.29 $\pm$ 0.11&4.5 $\pm$ 2.2\\
Present work& $^{13}$C($^{11}$B,$^{7}$Li)&50&0.37 $\pm$ 0.12&4.0 $\pm$ 1.1\\
\tableline
\end{tabular}
\tablenotetext{a}{Reanalysis of the experimental data of
\citet{kub03}.} \tablenotetext{b}{See Table 3 of \citet{kee03},
these values also depend on different normalization procedures used
in their work.}
\end{center}
\end{table}

\begin{table}
\begin{center}
\caption{Resonant parameters for $^{17}$O used in the present
calculations. $E_x$ and $E_R$ represent level energy and resonant
energy, respectively. $E_x$, $E_R$, $J^{\pi}$, $\Gamma_n(E_{R})$ and
$\Gamma_\alpha(E_{R})$ are taken from the compilation by
\citet{til93}. \label{tab3}}
\begin{tabular}{ccccccc}
\tableline\tableline
Level&$E_x$ (keV) & $E_R$ (keV) & $J^{\pi}$ & $\Gamma_n(E_{R})$ (keV) &$\Gamma_\alpha(E_{R})$ (keV)&$\gamma_\alpha^2$ (keV) \\
\tableline 1&6356 & $-$3 &1/2$^+$ &124 &&12.7\tablenotemark{a}\\
2&7165&806&5/2$^-$&1.38&0.0033&\\
3&7201&842&3/2$^+$&340\tablenotemark{b}&0.08\tablenotemark{b}&\\
4&7378&1019&5/2$^+$&0.64&0.01&\\
5&7381&1022&5/2$^-$&0.96&0.003&\\
6&7558&1199&3/2$^-$&500&0.08&\\
7&7688&1329&7/2$^-$&13&0.01&\\
8&7956&1597&1/2$^+$&84&6.7&\\
9&7991&1632&1/2$^-$&250&16&\\
10&8058&1699&3/2$^+$&71&15&\\
11&8200&1841&3/2$^-$&48&4.0&\\
12&8342&1983&1/2$^+$&8.1&2.2&\\
\tableline
\end{tabular}
\tablenotetext{a}{Reduced $\alpha$-width from this work.}
\tablenotetext{b}{The $\alpha$- and neutron-widths for the 3/2$^+$
resonance at $E_R$ = 842 keV were adjusted to provide the best
fitting of the experimental data, which resulted in broader partial
widths than the recommended values [$\Gamma_{n}(E_{R})$ = 280 keV,
$\Gamma_{\alpha}(E_{R})$ = 0.07 keV] of \citet{til93}.}
\end{center}
\end{table}

\begin{table}
\begin{center}
\caption{\label{tab4} Numerical values of the present
$^{13}$C($\alpha$,\,$n$)$^{16}$O rates (cm$^3$/mol/s) with the
adopted value, upper and lower limits for the 0.04$-$3.0 GK
temperature range and the coefficients $a_i$ in Eq. \ref{eq8} for
these three rates. The overall fitting errors are all less than 7\%
at temperatures from 0.04 to 10.0 GK.}
\begin{tabular}{cccccccc}
\hline\hline &\multicolumn{3}{c}{Reaction rate}&&\multicolumn{3}{c}{Coefficients}\\
\cline{2-4}\cline{6-8}
$T_9$&Adopt & Upper& Lower&$a_i$ & Adopt & Upper& Lower\\
\hline
0.04&2.63$\times10^{-24}$&3.41$\times10^{-24}$&1.92$\times10^{-24}$&$a_{1}$ & 79.3008& 52.8016& 87.5453\\
0.05&1.63$\times10^{-21}$&2.07$\times10^{-21}$&1.23$\times10^{-21}$&$a_{2}$ & $-$0.304890 & $-$0.204024& $-$0.301248\\
0.06&2.17$\times10^{-19}$&2.70$\times10^{-19}$&1.67$\times10^{-19}$&$a_{3}$ & 7.43132& $-$23.2591& 9.17396\\
0.07&1.07$\times10^{-17}$&1.31$\times10^{-17}$&8.44$\times10^{-18}$&$a_{4}$ & $-$84.8689& $-$42.4710& $-$95.7947\\
0.08&2.67$\times10^{-16}$&3.21$\times10^{-16}$&2.14$\times10^{-16}$&$a_{5}$ & 3.65083& 35.1371& 4.66751\\
0.09&4.04$\times10^{-15}$&4.78$\times10^{-15}$&3.28$\times10^{-15}$&$a_{6}$ & $-$0.148015& $-$17.3173& $-$0.221941\\
0.10&4.19$\times10^{-14}$&4.90$\times10^{-14}$&3.46$\times10^{-14}$&$a_{7}$ & 37.6008& 6.45708& 40.9578\\
0.15&1.67$\times10^{-10}$&1.86$\times10^{-10}$&1.46$\times10^{-10}$&$a_{8}$ & 62.5775& 64.3536& 63.1390\\
0.20&3.33$\times10^{-8}$&3.59$\times10^{-8}$&3.03$\times10^{-8}$&$a_{9}$ & $-$0.0277331& $-$0.302435& 0.0195985\\
0.30&3.09$\times10^{-5}$&3.20$\times10^{-5}$&2.96$\times10^{-5}$&$a_{10}$ & $-$32.3917& 3.36966& $-$34.5026\\
0.50&8.06$\times10^{-2}$&8.12$\times10^{-2}$&7.98$\times10^{-2}$&$a_{11}$ & $-$48.9340& $-$64.0633& $-$47.2196\\
1.0 &2.44$\times10^{2}$  &2.45$\times10^{2}$  &2.44$\times10^{2}$  &$a_{12}$ & 44.1843& 1.62313& 44.0189\\
2.0 &5.99$\times10^{4}$  &6.00$\times10^{4}$  &5.97$\times10^{4}$  &$a_{13}$ & $-$20.8743& 0.00566612& $-$20.9558\\
3.0 &6.12$\times10^{5}$  &6.13$\times10^{5}$  &6.10$\times10^{5}$  &$a_{14}$ & 2.02494& 31.1730& 0.905521\\
\hline
\end{tabular}
\end{center}
\end{table}

\begin{table}
\begin{center}
\caption{Summary of the stellar models. \label{tab5}}
\begin{tabular}{ccc}
\tableline\tableline
Mass ($M_\sun$) & Metallicity ($Z$) & Number of $^{13}$C pockets \\
\tableline
1.0 & 0.0001 &2\\
1.5 & 0.0001 &15\\
1.8 & 0.01 &6\\
3.0 & 0.02 &16\\
\tableline
\end{tabular}
\end{center}
\end{table}

\end{CJK*}
\end{document}